\documentclass[12pt,preprint,usenatbib]{aastex}

\shorttitle{How binaries affect stellar population synthesis}

\shortauthors{Li \& Han}

\begin{document}
\title{How binary interactions affect spectral stellar population synthesis}
\author{Zhongmu Li\footnote{Graduate University of the Chinese Academy of Sciences} ~and Zhanwen Han}

\affil{National Astronomical Observatories/Yunnan Observatory, the
Chinese Academy of Sciences, Kunming, 650011,
   China}
\email{zhongmu.li@gmail.com; zhanwenhan@hotmail.com}

\begin{abstract}

Single-star stellar population (ssSSP) models are usually used for
spectral stellar population studies. However, more than 50\% of
stars are in binaries and evolve differently from single stars. This
suggests that the effects of binary interactions should be
considered when modeling the stellar populations of galaxies and
star clusters. Via a rapid spectral stellar population synthesis
($RPS$) model, we give detailed studies of the effects of binary
interactions on the Lick indices and colours of stellar populations,
and on the determination of the stellar ages and metallicities of
populations.

Our results show that binary interactions make stellar populations
less luminous, bluer, with larger age-sensitive Lick index
(H$\beta$) and smaller metallicity-sensitive indices (e.g., Mgb,
Fe5270 and Fe5335) compared to ssSSPs. It also shows that when ssSSP
models are used to determine the ages and metallicities of stellar
populations, smaller ages or metallicities will be obtained, when
using two line indices (H$\beta$ and [MgFe]) and two colours (e.g.,
$u-R$ and $R-K$), respectively. Some relations for linking the
stellar-population parameters obtained by ssSSPs to those obtained
by binary-star stellar populations (bsSSPs) are presented in the
work. This can help us to get some absolute values for
stellar-population parameters and is useful for absolute studies.
However, it is found that the relative luminosity-weighted stellar
ages and metallicities obtained via ssSSPs and bsSSPs are similar.
This suggests that ssSSPs can be used for most spectral stellar
population studies, except in some special cases.

\end{abstract}


\section{Introduction}

Stellar population synthesis is a powerful technique to study the
stellar contents of galaxies and star clusters (see, e.g.,
\citealt{Yungelson:1997}, \citealt{Tout:1997}, \citealt{Pols:1998},
\citealt{Hurley:2007}). It is also an important method to study the
formation and evolution of galaxies. Simple stellar population (SSP)
models that do not take binary interactions into account are usually
used for spectral stellar population studies, as most models are
ssSSP models (e.g., \citealt{Bruzual:2003}, \citealt{Vazdekis:1999},
\citealt{Fioc:1997}, \citealt{Worthey:1994}). However, as pointed
by, e.g., \cite{Duquennoy:1991}, \cite{Pinfield:2003}, and
\cite{Lodieu:2007}, about 50\% of stars are in binaries and they
evolve differently from single stars. We can see this when comparing
the isochrone of an ssSSP to that of a bsSSP (Fig.1). In fact,
bsSSPs better fit the colour-magnitude diagrams (CMDs) of star
clusters than ssSSPs \citep{Li:2007database}. This suggests that
binary interactions can affect stellar population synthesis studies
significantly and it is, therefore, necessary to consider binary
interactions. This is also supported by some observational results,
e.g., the Far-UV excess of elliptical galaxies \citep{Han:2007} and
blue stragglers in star clusters (e.g., \citealt{Davies:2004};
\citealt{Tian:2006}; \citealt{Xin:2007}). These phenomena can be
naturally explained via stellar populations with binary
interactions, without any special assumptions.

A few works have tried to model populations via binary stars and
have presented some results on the effects of binary interactions on
spectral stellar population synthesis. For example,
\citet{Zhang:2004, Zhang:2005} showed that binary interactions can
make bsSSPs bluer than ssSSPs. However, there is not a more detailed
investigation about how binary interactions affect the Lick indices
and colours of stellar populations, and on the determination of
stellar ages and metallicities. One of our previous works,
\citet{Li:2007database}, compared bsSSPs to ssSSPs, but various
stellar population models are used. This makes it difficult to
understand the effects of the changes of Lick indices and colours of
populations that result only from binary interactions. Furthermore,
it did not show how binary interactions affect the determinations of
stellar ages and metallicities. In this case, we have no clear
picture for the differences between the predictions of bsSSPs and
ssSSPs, and do not well know the differences between
luminosity-weighted stellar-population parameters (age and
metallicity) determined by bsSSPs and ssSSPs. Because all galaxies
contain some binaries, detailed studies of the effects of binary
interactions on the Lick indices and colours of populations are
important, as are the determinations of stellar-population
parameters. In this work, we perform a detailed study of the effects
of binary interactions on spectral stellar population synthesis
studies, via the rapid spectral population synthesis ($RPS$) model
of \citet{Li:2007database}.

The paper is organized as follows. In Sect. 2, we briefly introduce
the $RPS$ model. In Sect. 3, we study the effects of binary
interactions on the isochrones, spectral energy distributions
(SEDs), Lick Observatory Image Dissector Scanner absorption line
indices (Lick indices), and colours of stellar populations. In Sect.
4, we investigate the differences between stellar ages and
metallicities fitted by ssSSPs and bsSSPs. Finally, in Sect. 5, we
give our discussions and conclusions.

\section{The rapid spectral population synthesis model}

We take the results of the $RPS$ model of \citet{Li:2007database}
for the work, as there is no other available model. The $RPS$ model
calculated the evolution of binaries and single stars via the rapid
stellar evolution code of \citet{Hurley:2002} (hereafter Hurley
code) and took the spectral libraries of \citet{Martins:2005} and
\citet{Westera:2002} (BaSeL 3.1) for spectral synthesis. The model
calculated the high-resolution (0.3 $\rm \AA$) SEDs, Lick indices
and colours for both bsSSPs and ssSSPs with the initial mass
functions (IMFs) of \citet{Salpeter:1955} and \citet{Chabrier:2003}.
Note that the $RPS$ model used a statistical isochrone database for
modeling stellar populations \citep{Li:2007database}. Each bsSSP
contains about 50\% stars that are in binaries with orbital periods
less than 100 yr (the typical value of the Galaxy), and binary
interactions such as mass transfer, mass accretion, common-envelope
evolution, collisions, supernova kicks, angular momentum loss
mechanism, and tidal interactions are considered when evolving
binaries via Hurley code. Thus the $RPS$ model is suitable for
studying the effects of binary interactions on stellar population
synthesis studies. However, some parameters such as the ones used
for describing the common envelope prescription, mass-loss rates,
and supernova kicks are free parameters and the default values in
Hurley code, i.e., 0.5, 1.5, 1.0, 0.0, 0.001, 3.0, 190.0, 0.5, and
0.5, are taken in this work for wind velocity factor ($\beta_{\rm
w}$), Bondi-Hoyle wind accretion faction ($\alpha_{\rm w}$), wind
accretion efficiency factor ($\mu_{\rm w}$), binary enhanced mass
loss parameter ($B_{\rm w}$), fraction of accreted material retained
in supernova eruption ($\epsilon$), common-envelope efficiency
($\alpha_{\rm CE}$), dispersion in the Maxwellian distribution for
the supernovae kick speed ($\sigma_{\rm k}$), Reimers coefficient
for mass loss ($\eta$), and binding energy factor ($\lambda$),
respectively. These default values are taken because they have been
tested by the developer of Hurley code and seem more reliable. One
can refer to the paper of \citet{Hurley:2002} for more details. In
fact, many of these free parameters remain uncertain, and their
uncertainties can possibly have great effect on our results. When we
test the uncertainties caused by various $\alpha_{\rm CE}$ and
$\lambda$, we find that the number of blue stragglers can be changed
as large as 40\% compared to the default case. However, it is
extremely difficult to give detailed uncertainties in spectral
stellar population synthesis due to the free parameters, as we lack
of constrains on these free parameters (see \citealt{Hurley:2002}).
Therefore, when we estimate the synthetic uncertainties in our $RPS$
model, the uncertainties due to variation of free parameters will
not be taken into account. Because the fitted formulae used by
Hurley code to evolve stars lead to uncertainties less than about
5\% \citet{Hurley:2002}, we take 5\% as the uncertainties in the
evolution of stars in the whole paper. The correctness of the
results of the $RPS$ model depends on how correct the default
parameters of Hurley code are. In addition, the uncertainties in
final generated spectrum caused by the spectral library and the
method used for spectral synthesis are about 3\% and 0.81\%,
respectively. Because a Monte Carlo technique is used by the $RPS$
model to generate the star sample (2\,000\,000 binaries or
4\,000\,000 single stars in our work and it is twice of that of the
model of \citealt{Zhang:2004}) of stellar populations, the number of
stars can result in statistical errors in the Lick indices and
colours of populations. According to our test, 1\,000\,000 binaries
is enough to get reliable Lick indices (see also
\citealt{Zhang:2005}), but the near-infrared colours such as
$(I-K)$, $(R-K)$, and $(r-K)$ of old populations are affected by the
Monte Carlo method. However, the errors caused by the Monte Carlo
method are small, about 2\% for a sample of 4\,000\,000 stars. Note
that in this work, an uniform distribution is used to generate the
ratio ($q$, 0--1) of the mass of the secondary to that of the
primary (\citealt{Mazeh:1992}; \citealt{Goldberg:1994}), and then
the mass of the secondary is calculated from that of the primary and
$q$. The separation ($a$) of two components of a binary is generated
following the assumption that the fraction of binary in an interval
of log($a$) is constant when $a$ is big (10$R_\odot$ $< a <$ 5.75
$\times$ 10$^{\rm 6}$$R_\odot$) and it falls off smoothly when when
$a$ is small ($\leq$ 10$R_\odot$) \citep{Han:1995}. The distribution
of $a$ is written as
\begin{equation}
  a~.p(a) = \left\{
            \begin{array}{ll}
            a_{\rm sep}(a/a_{\rm 0})^{\psi}, &~a \leq a_{\rm 0}\\
            a_{\rm sep}, &~a_{\rm 0} < a < a_{\rm 1}\\
     \end{array}
    \right.
\end{equation}
where $a_{\rm sep} \approx 0.070, a_{\rm 0} = 10R_{\odot}, a_{\rm 1}
= 5.75 \times 10^{\rm 6}R_\odot$ and $\psi \approx 1.2$. The
eccentricity ($e$) of each binary system using a uniform
distribution, in the range of 0--1, and $e$ affects the results
slightly \citep{Hurley:2002}.

In addition, $RPS$ model uses some methods different from those used
in the work of \citet{Zhang:2004} to calculate the SEDs, Lick
indices, and colours of populations. Besides $RPS$ used a
statistical isochrone database, it calculated the Lick indices
directly from SEDs while the work of \citet{Zhang:2004} used some
fitting formulae to compute the same indices. The $RPS$ model
calculated the colours of populations from SEDs, but the work of
\citet{Zhang:2004} computed colours by interpolating the photometry
library of BaSeL 2.2 \citep{Lejeune:1998}. Furthermore, the $RPS$
model used the more advanced version 3.1 \citep{Westera:2002} of
BaSeL library rather than version 2.2 to give the colours of the
populations. The BaSeL 3.1 library overcomes the weakness of the
BaSeL 2.2 library at low metallicity, because it has been
colour-calibrated independently at all levels of metallicity. This
makes the predictions of our model are more reliable. Another
important point is that the model of Zhang et al. did not present
the near-infrared colours of stellar populations, but such colours
are very important for disentangling the well-known age--metallicity
degeneracy.

\section{Effects of binary interactions on stellar population synthesis}

\subsection{The effects on isochrones of stellar populations}
The direct effect of binary interactions on stellar population
synthesis is to change the isochrones of stellar populations, e.g.,
the distribution of stars in the surface gravity [log($g$)] versus
effective temperature ($T_{\rm eff}$) grid, hereafter $gT$-grid. We
investigate the differences between the isochrones of bsSSPs and
ssSSPs. Stellar populations with the IMF of \citet{Salpeter:1955}
are taken as our standard models for this work. The Salpeter IMF is
actually not the best one for stellar population studies, although
it is widely used. The reason is that the IMF is not valid for low
masses. However, this IMF is reliable for stellar population
synthesis, because low-mass stars contribute much less to the light
of populations compared to high-mass stars. Some further studies
will be given by taking some more reliable IMFs, e.g.,
\citet{Kroupa:1993}. Because the isochrone database used by this
work divides the $gT$-grid into 1\,089\,701 sub-grids, with 0.01 and
40K the intervals of log($g$) and $T_{\rm eff}$, it is possible to
compare the isochrones of bsSSPs and ssSSPs. The differences between
the isochrones of two kinds of populations are calculated by
subtracting the fraction of stars of bsSSP from that of its
corresponding ssSSP, sub-grid by sub-grid. The ssSSP and its
corresponding bsSSP have the same star sample, metallicity and age,
and all their integrated specialties (SEDs, colours and Lick
indices) are calculated via the same method. Therefore, the
differences between the isochrones of a bsSSP and its corresponding
ssSSP only result from binary interactions. For convenience, we call
the difference a ``discrepancy isochrone''. Here we show the
discrepancy isochrones for a few stellar populations in Figs. 2 and
3, for metal-poor ($Z$ = 0.004) and solar-metallicity ($Z$ = 0.02)
populations, respectively. Because it is found that the discrepancy
isochrones of metal-rich ($Z$ = 0.03) populations are similar to
those of solar-metallicity populations, we do not show the results
for metal-rich populations. Note that the results for populations
with metallicities poorer than 0.004 are also given by our work, but
we do not show them as the example for metal-poor populations as the
$RPS$ model did not give the SEDs and Lick indices for populations
with metallicities poorer than 0.004. This is actually limited by
the spectral library used by the $RPS$ model, which only supplies
spectra for stars more metal rich than 0.002. As we see, some
special stars, e.g., blue stragglers, are generated by binary
interactions (see also Fig. 1). We also show that the differences
between isochrones of old bsSSPs and ssSSPs are smaller than those
of young populations, because the isochrones of old populations are
dominated by low-mass stars, in which binary interactions are much
weaker.
\begin{figure} 
\includegraphics[angle=-90,width=160mm]{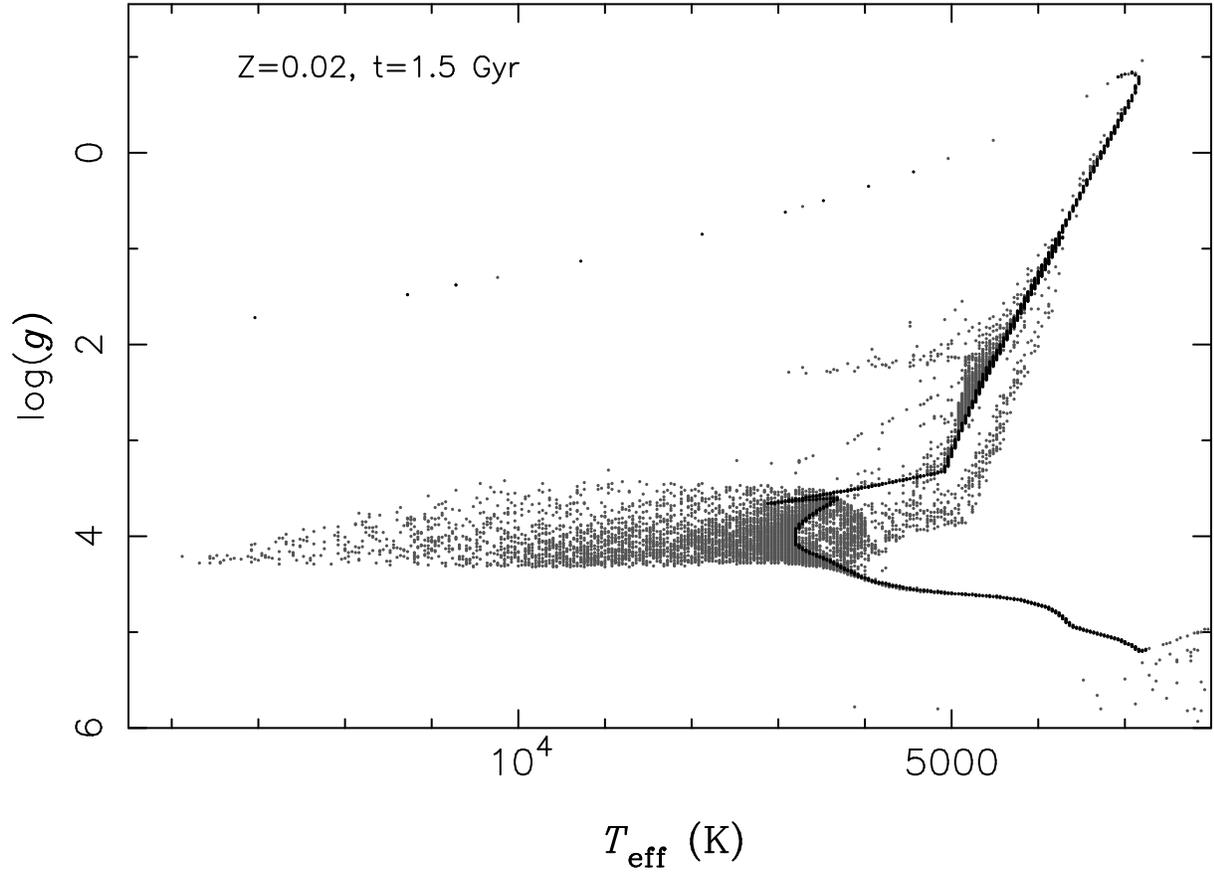}
\caption{Comparison of the shapes of the isochrones of a pair of
solar-metallicity ($Z$ = 0.02) bsSSP and ssSSP. The figure is
plotted by putting the isochrones of the bsSSP and ssSSP together.
Black points show the isochrone of the ssSSP, and gray points the
isochrone of the bsSSP.}
\end{figure}

\begin{figure} 
\includegraphics[angle=-90,width=160mm]{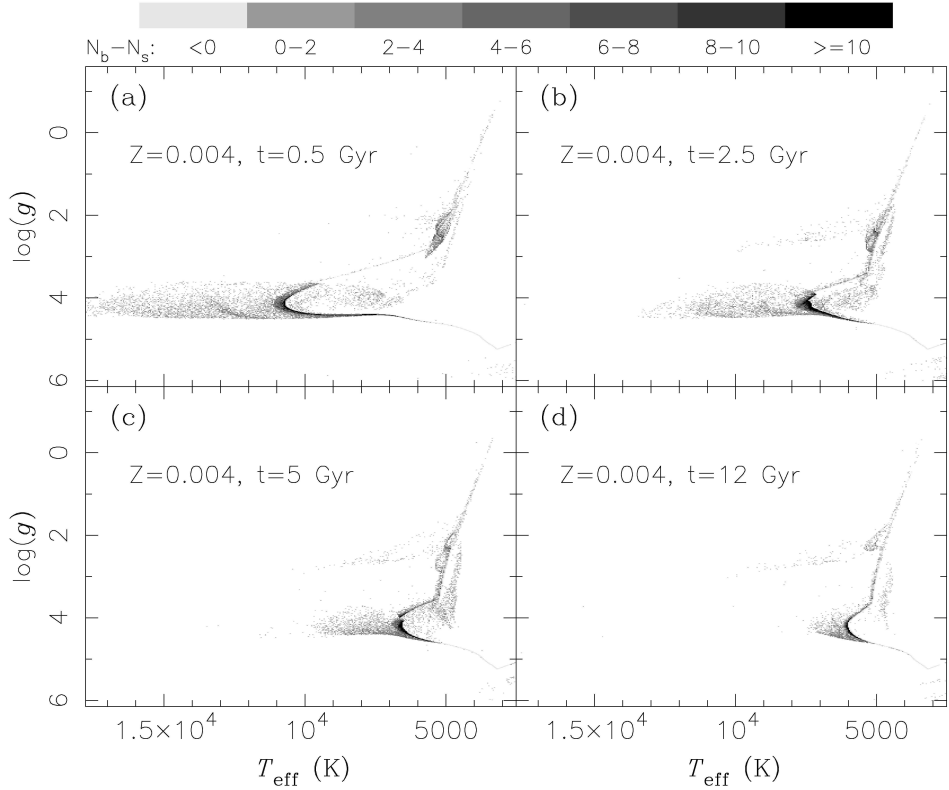}
\caption{Differences between the number distribution of stars in the
isochrones of metal-poor ($Z$ = 0.004) bsSSPs and ssSSPs.The darker
the colour, the bigger the difference between star numbers of the
bsSSP and ssSSP, (N$_{\rm b}$ - N$_{\rm s}$). For each sub-grid,
N$_{\rm b}$ is the number of stars in a bsSSP, while N$_{\rm s}$ the
number of stars in an ssSSP and locating in the sub-grid. Note that
the grid of log($g$) versus $T_{\rm eff}$ with log($g$) range of
-1.5 -- 6 and $T_{\rm eff}$ range of 2\,000 -- 60\,000 K is divided
into 1\,089\,701 sub-grids, with intervals of 0.01 and 40 K for
log($g$) and $T_{\rm eff}$, respectively. A point in the figures
corresponds to a sub-grid.}
\end{figure}

\begin{figure} 
\includegraphics[angle=-90,width=160mm]{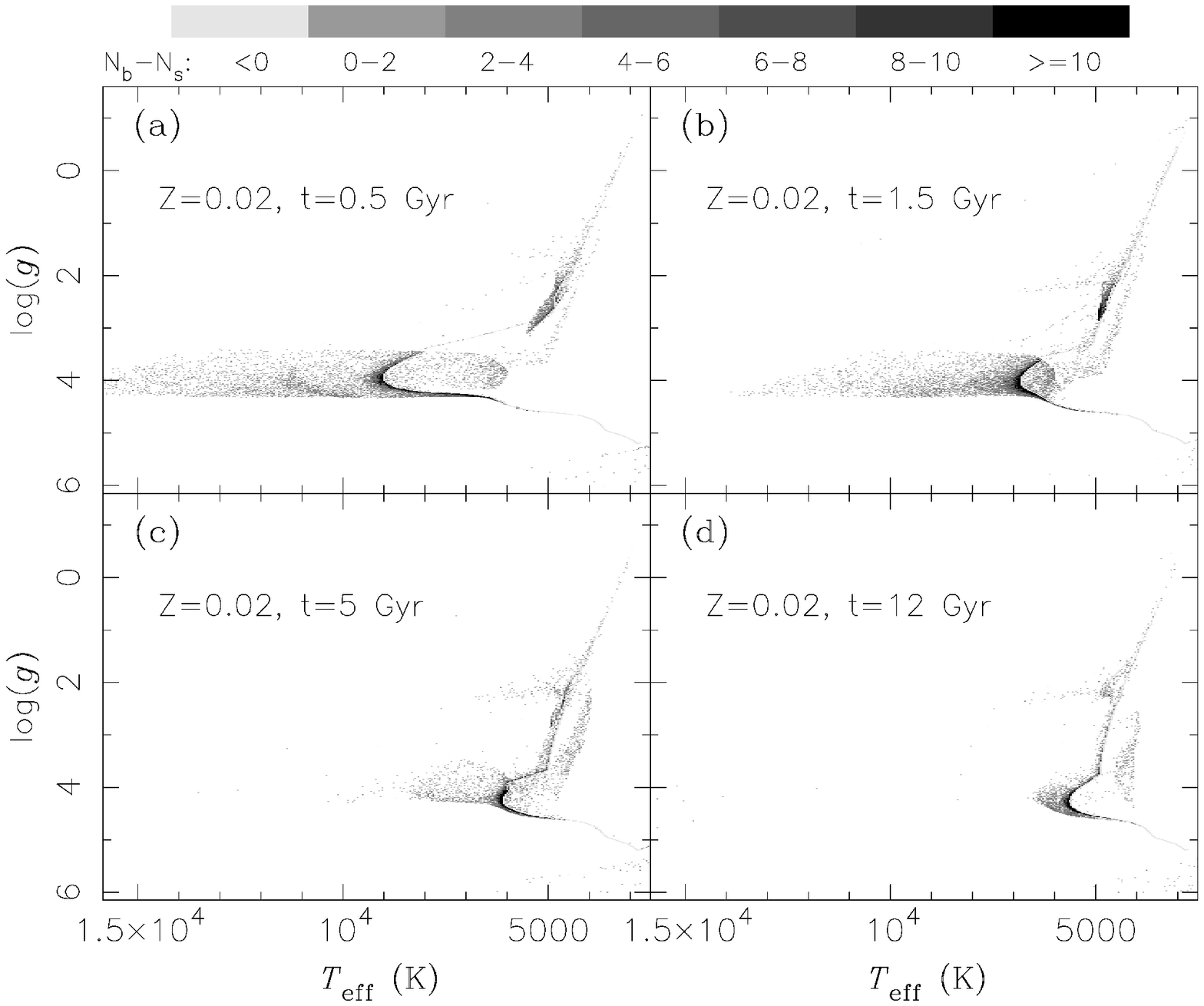}
\caption{Similar to Fig.2, but for solar-metallicity ($Z$ = 0.02)
stellar populations.}
\end{figure}

\subsection{The effects on integrated features of populations}

The widely used integrated features of stellar populations are SEDs,
Lick indices and colours. They are usually used for stellar
population studies and are important. We investigate the effects of
binary interactions on them in this section.

\subsubsection{Spectral energy distributions}

To investigate how binary interactions affect the SEDs of stellar
populations, we compare the SEDs of a bsSSP and an ssSSP that have
the same age and metallicity. The difference between SEDs are simply
called discrepancy SEDs. The absolute discrepancy SED for a pair of
bsSSP and ssSSP is derived by subtracting the flux of the ssSSP from
that of the bsSSP, as a function of wavelength. The discrepancy SEDs
are mainly caused by blue stragglers and hot sub-dwarfs, as such
stars are very hot and luminous. The changes of surface abundances
of stars caused by binary interactions can also contribute to
discrepancy SEDs. The absolute discrepancy SEDs for metal-poor ($Z$
= 0.004), and solar-metallicity ($Z$ = 0.02) stellar populations are
shown in Figs. 4, and 5, respectively. The absolute discrepancy SEDs
such as those shown in the two figures can be easily used to add
binary interactions into ssSSP models, but fractional discrepancy
SEDs are more useful for understanding the effects of binary
interactions. We show the fractional discrepancy SEDs of a few
solar-metallicity populations in Fig. 6.
\begin{figure} 
\includegraphics[angle=-90,width=160mm]{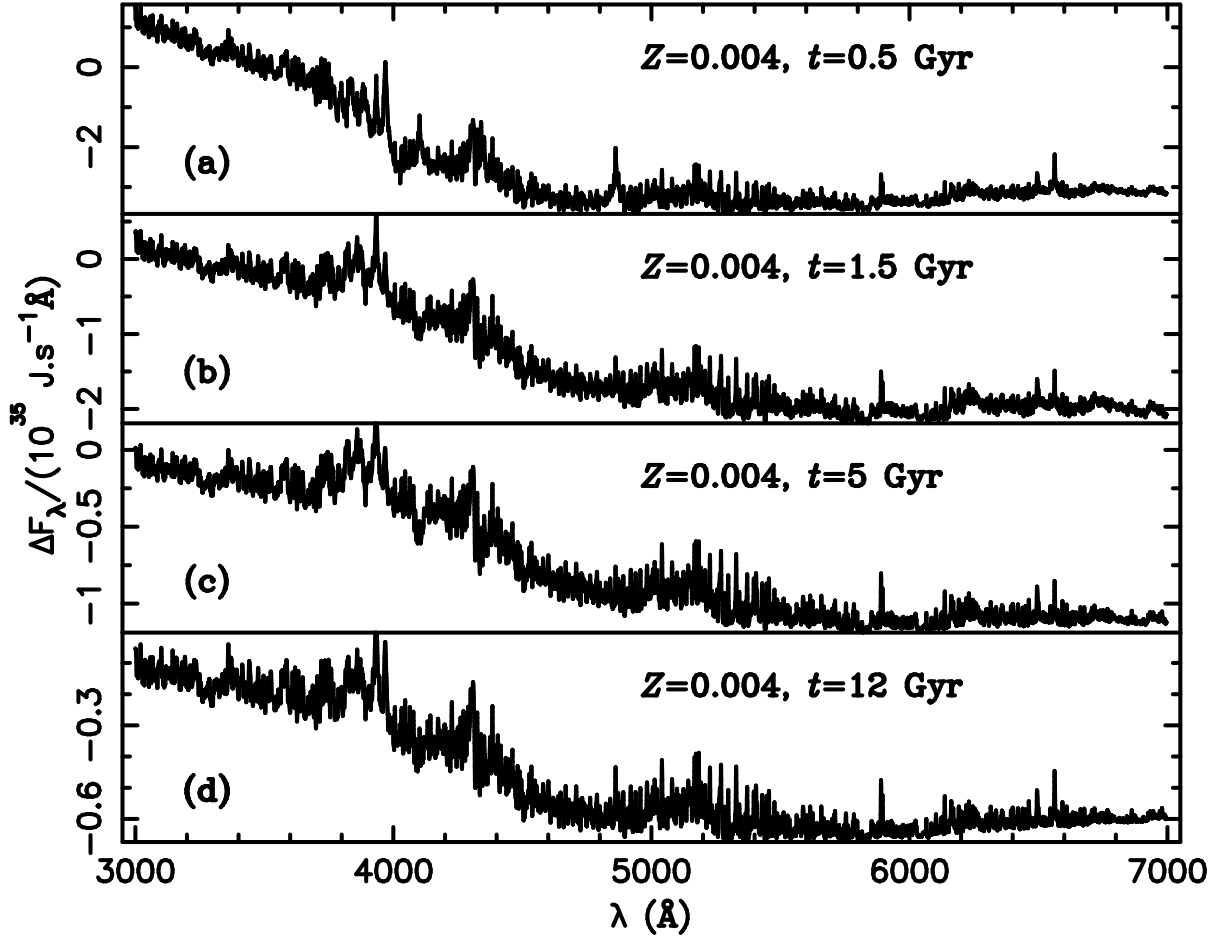}
\caption{Absolute differences between the SEDs of metal-poor ($Z$ =
0.004) bsSSPs and ssSSPs, i.e., absolute discrepancy SEDs. The
discrepancy SED of a pair of bsSSP and ssSSP is calculated by
subtracting the flux of the ssSSP, F$_{\rm s}$, from that of the
bsSSP, F$_{\rm b}$, as a function of wavelength. Each population
contains 2\,000\,000 binaries or 4\,000\,000 single stars, whose
initial mass follows the Salpeter IMF and with the lower and upper
limits of 0.1 and 100 $M_{\bigodot}$.}
\end{figure}

\begin{figure} 
\includegraphics[angle=-90,width=160mm]{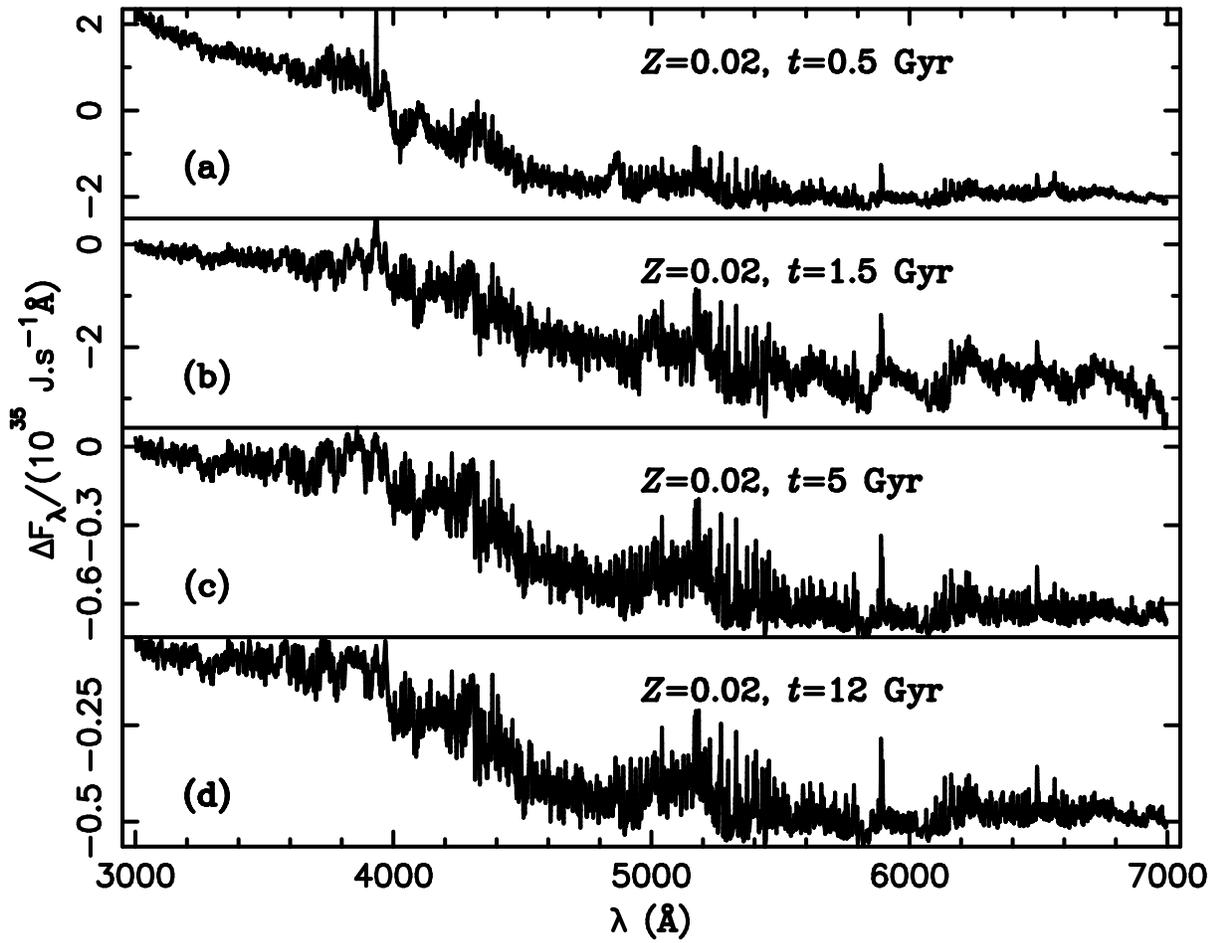}
\caption{Similar to Fig. 4, but for solar-metallicity ($Z$ = 0.02)
stellar populations.}
\end{figure}

\begin{figure} 
\includegraphics[angle=-90,width=160mm]{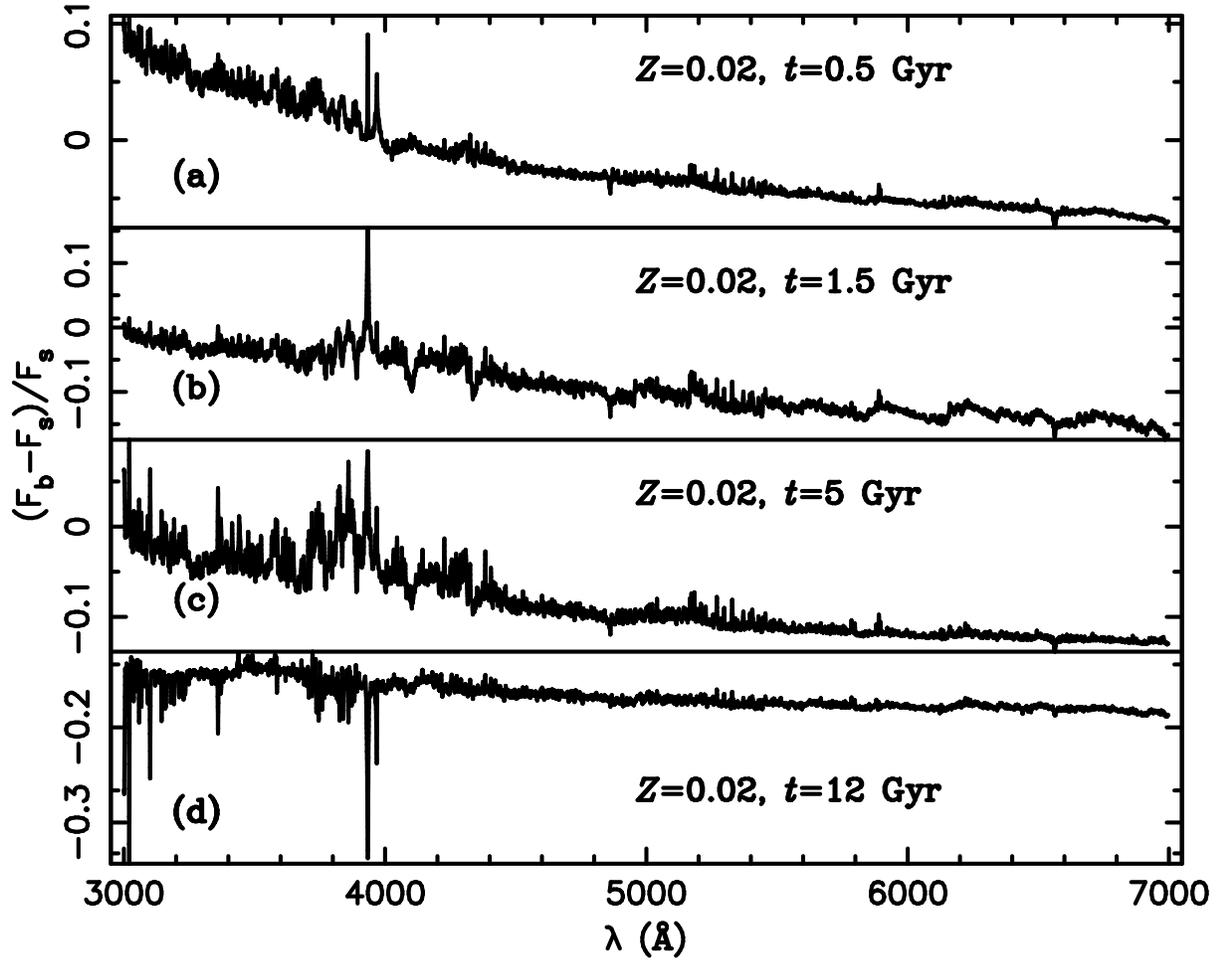}
\caption{Fractional discrepancy SEDs of solar-metallicity stellar
populations. Symbols F$_{\rm s}$ and F$_{\rm b}$ have the same
meanings as in Fig. 4.}
\end{figure}

As we see, binary interactions make stellar populations less
luminous, but the flux in shortwave bands are changed by binary
interactions weakly compared to that in longwave bands. This mainly
results from special stars generated by binary interactions, which
contribute differently to flux in different bands. The differences
between the SEDs of a bsSSP and its corresponding ssSSP decrease
with increasing age or decreasing metallicity. In addition, it
suggests that binary interactions can affect most Lick indices and
colours of populations, because the flux changes caused by binary
interactions are not 0, in the bands where widely used Lick indices
and magnitudes are defined. In this case, bsSSP and ssSSP models
usually give different results for stellar population studies.
Furthermore, because the effects of binary interactions on the SED
flux of populations are about 11\% on average, they are detectable
for observations with spectral signal to noise ratio (SNR) grater
than 10. In other words, the effects can be detected by most
observations, as most reliable observations have SNRs greater than
10.

\subsubsection{Lick indices}

Lick indices are the most widely used indices in stellar population
studies, because they can disentangle the well-known stellar
age--metallicity degeneracy (see, e.g., \citealt{Worthey:1994}). If
binary interactions are taken into account, some results different
from those determined via ssSSP models will be obtained, which was
suggested by the study of the differences between the SEDs of bsSSPs
and ssSSPs. Here we test how binary interactions change the Lick
indices of stellar populations when comparing to those of ssSSPs. In
Fig. 7, we show the differences between four widely used indices of
bsSSPs and those of ssSSPs.
\begin{figure} 
\includegraphics[angle=-90,width=160mm]{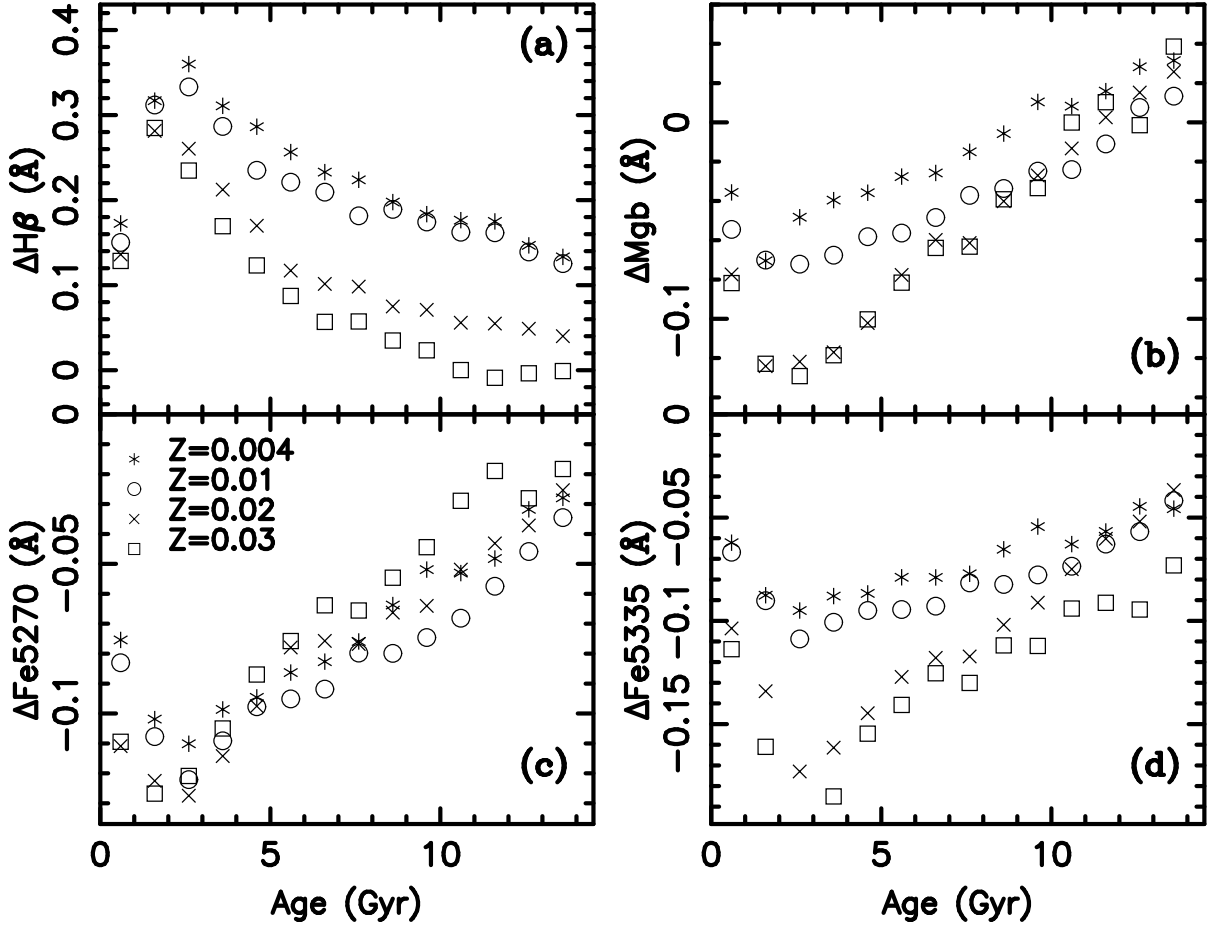}
\caption{Differences between four widely used Lick indices of bsSSPs
and ssSSPs. The difference in a Lick index is calculated by
subtracting the value of a ssSSP from that of its corresponding
bsSSP, which has the same age and metallicity as the ssSSP. Star,
circle, cross, and square are for populations with metallicities of
0.004, 0.01, 0.02, and 0.03, respectively. Note that the differences
are averaged in each bin.}
\end{figure}

The indices are calculated from SEDs on the Lick system
(\citealt{Worthey:1994lickdefinition}) directly. As we see, Fig. 7
shows that binary interactions make the H$\beta$ index of a
population larger by about 0.15 ${\rm \AA}$, while making Mgb index
smaller by about 0.06 ${\rm \AA}$ and Fe indices smaller by more
than about 0.1 ${\rm \AA}$, compared to ssSSPs. Therefore, the
changes in Lick indices are usually larger than typical
observational uncertainties (about 0.07 ${\rm \AA}$ for H$\beta$
index and 0.04 ${\rm \AA}$ for metal-line indices according to the
data of \citealt{Thomas:2005}). For fixed metallicity, the effects
of binary interactions on both age- and metallicity-sensitive
indices become stronger with increasing age when stellar age is less
than about 1.5 $\sim$ 2\,Gyr, and then the effects become weaker
with increasing age. The reason is that binary interactions change
the isochrones of populations most strongly near 1.5 $\sim$ 2\,Gyr
as the first mass transfer between two components of binaries peak
and a lot of blue stragglers are generated near 1.5 $\sim$ 2\,Gyr
according to the star sample of bsSSPs, and the light of old
populations is dominated by low-mass binaries. The interactions
between two components of low-mass binaries are usually weaker than
for high-mass binaries. The effects of binary interactions on the
isochrones are tested quantitatively using the numbers of stars with
log($g$) $<$ 4.0 and log($T_{\rm eff}$) $>$ 3.75, because these
stars are very luminous and contribute a lot to the light of their
populations. Our result shows that binary interactions change the
number distribution of stars in the above log($g$) and log($T_{\rm
eff}$) ranges most significantly when stellar age is from 1.5 to 2.
In addition, from Fig. 7, we find that for a fixed age, binary
interactions affect the H$\beta$ and Fe5270 indices of metal-poor
populations more strongly than those of metal-rich populations,
while they affect the Mgb and Fe5335 indices of metal-poor
populations more weakly. As a whole, using ssSSPs and bsSSPs,
various ages and metallicities will be measured for the same stellar
population, via popular Lick-index methods such as the H$\beta$ \&
[MgFe] method, which determines the ages and metallicities of
populations by comparing the observational and theoretical H$\beta$
and [MgFe] indices \citep{Thomas:2003}. Note that the evolution of
the differences between the Lick indices of ssSSPs and bsSSPs were
not shown before.

\subsubsection{Colour indices}

Because colours are useful for estimating the ages and metallicities
of distant galaxies (see, e.g., \citealt{Li:2008colourpairs}), we
investigate the effects of binary interactions on them. We use a
method similar to that used for studying the Lick indices of stellar
populations. Our detailed results are shown in Fig. 8.
\begin{figure} 
\includegraphics[angle=-90,width=160mm]{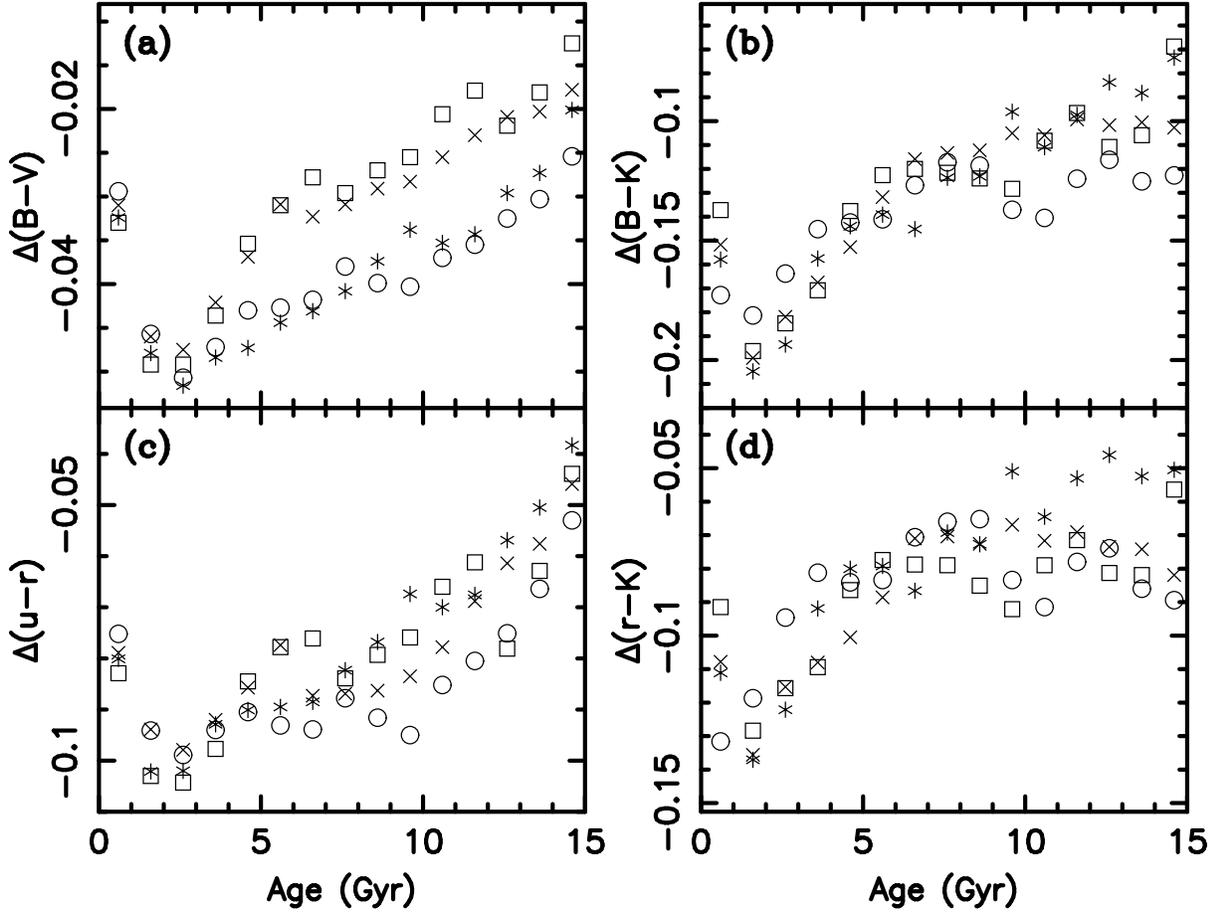}
\caption{Differences between four colours of bsSSPs and ssSSPs. The
differences in a colour are calculated by subtracting the colour of
a ssSSP from that of its corresponding bsSSP (with the same age and
metallicity). Symbols have the same meanings as in Fig. 6. The
colour $(u-r)$ is on the SDSS-$ugriz$ system and $(r-K)$ is a
composite colour that consists of a Johnson magnitude ($K$) and a
SDSS-$ugriz$ magnitude ($r$). The differences are averaged in each
bin.}
\end{figure}
In the figure, the differences between two $UBVRIJHK$ colours,
$(B-V)$ and $(B-K)$, a $ugriz$ colour on the photometric system used
by Sloan Digital Sky Survey (hereafter SDSS system), $(u-r)$, and a
composite colour, $(r-K)$, of bsSSPs and those of ssSSPs are shown,
respectively. Note that the $(r-K)$ colour consists of a Johnson
system magnitude, $K$, and an SDSS system magnitude, $r$. As we see,
for a fixed age and metallicity, binary interactions make the
colours of most stellar populations bluer than those predicted by
ssSSPs. This mainly results from blue stragglers generated by binary
interactions, because such stars are very luminous and blue. It
suggests that we will get different stellar metallicities and ages
for galaxies via bsSSP and ssSSP models, using a photometric method.
When comparing to the typical colour uncertainties (0.12, 0.06,
0.13, 0.10, and 0.01 mag for $B$, $V$, $K$, $u$, and $r$ magnitudes,
respectively), the changes [e.g., about -0.04, -0.15, and -0.08 mag
for $(B-V)$, $(B-K)$, and $(u-r)$, respectively] of colours caused
by binary interactions are similar to, but somewhat less than,
typical observational errors. Note that the photometric
uncertainties are estimated using the data of some publications,
SDSS, and Two Micron All Sky Survey (2MASS). The uncertainties
actually depend on surveys. The observational uncertainties of $u$
and $K$ magnitudes may be smaller when taking the data of other
surveys instead of those of SDSS and 2MASS. In addition, similar to
Lick indices, the differences between the colours of two kinds of
stellar populations peak near 2\,Gyr.

\section{The effects of binary interactions on the determination of
stellar-population parameters}

Two stellar-population parameters, i.e., stellar age and
metallicity, are crucial in the investigations of the formation and
evolution of galaxies. We investigate the effects of binary
interactions on the estimates of the two parameters. We try to fit
the stellar-population parameters of bsSSPs with various ages and
metallicities using ssSSPs, via both Lick-index and photometric
methods. Because observations show that about 50\% of stars in the
Galaxy are in binaries, bsSSPs should be more similar to the real
stellar populations of galaxies and star clusters. Therefore, the
stellar-population parameters fitted via ssSSPs (hereafter ss-fitted
results, represented by $t_{\rm s}$ and $Z_{\rm s}$) should be
different from the results obtained via bsSSPs (bs-fitted results,
$t_{\rm b}$ and $Z_{\rm b}$). The detailed differences are shown in
this section.

\subsection{Lick-index method}

In a widely used method, i.e., Lick-index method, we fit the stellar
ages and metallicities of populations by two indices, i.e., H$\beta$
and [MgFe] = $\rm {\sqrt{Mgb . (0.72Fe5270 + 0.28Fe5335)}}$, after
\citet{Thomas:2003}. Thus the results are slightly affected by
$\alpha$-enhancement and stellar population models (e.g., our $RPS$
model) without any $\alpha$-enhancement compared to the sun can be
used to measure stellar-population parameters. The differences
between bs- and ss-fitted stellar-populations parameters of
populations with four metallicities (0.004, 0.01, 0.02, and 0.03)
and 150 ages (from 0.1 to 15\,Gyr) are tested. In the test, we try
to fit the stellar ages and metallicities of testing bsSSPs via an
H$\beta$ versus [MgFe] grid of ssSSPs. Because ssSSPs predict
different Lick indices for populations compared to bsSSPs, when we
use ssSSPs to fit the ages and metallicities of our testing bsSSPs,
the results obtained are different from the real parameters of
bsSSPs, i.e., bs-fitted parameters. From an H$\beta$ versus [MgFe]
grid of ssSSPs (Fig. 9), we can see this clearly.
\begin{figure} 
\includegraphics[angle=-90,width=160mm]{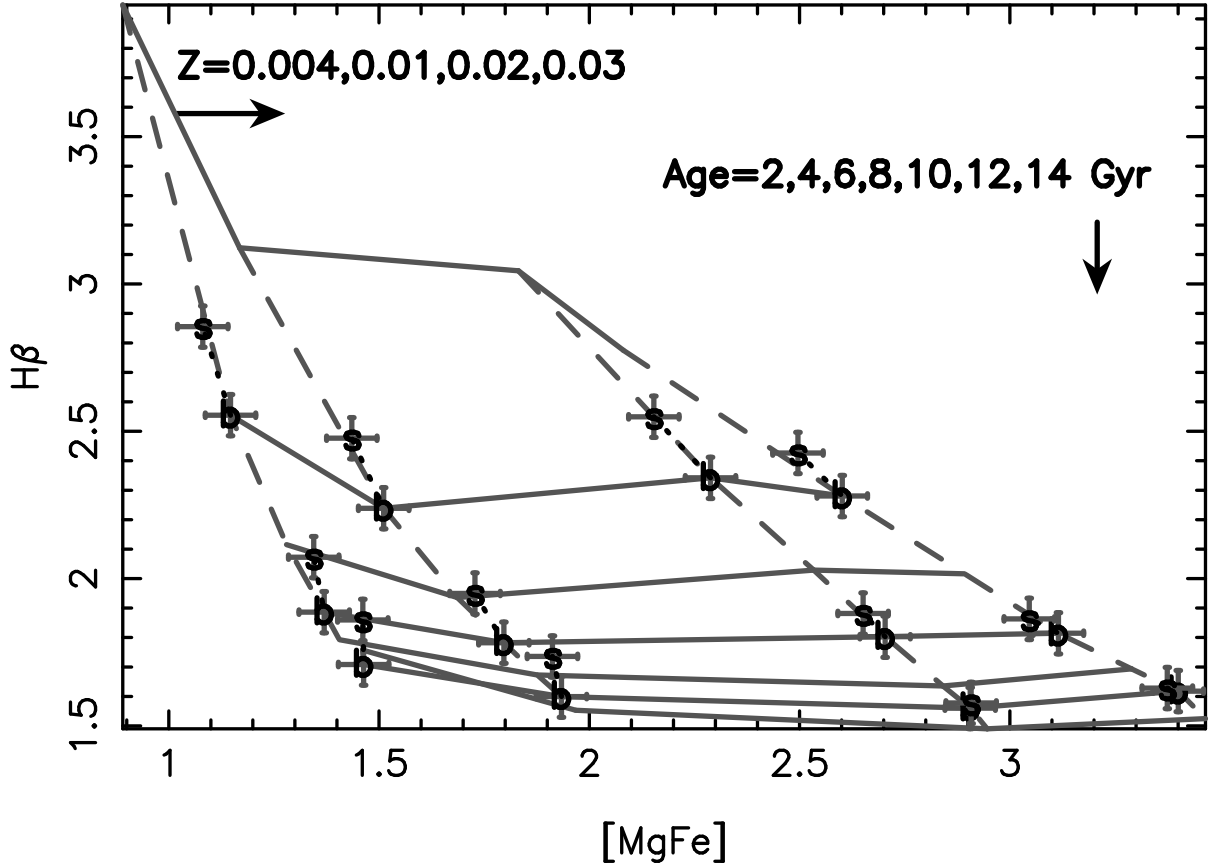}
\caption{Some bsSSPs on the H$\beta$ versus [MgFe] grid of ssSSPs.
The position ``b'' shows the bs-fitted age and metallicity of a
bsSSP, and ``s'' shows the ss-fitted values for it. The composite
index [MgFe] is calculated by [MgFe] = $\rm {\sqrt{Mgb . (0.72Fe5270
+ 0.28Fe5335)}}$ \citep{Thomas:2003}. Dashed and solid lines are for
constant metallicity and constant age, respectively. Error bars show
the typical observational uncertainties of indices.}
\end{figure}
In detail, the ss-fitted stellar ages are less than the bs-fitted
ones, by 0 $\sim$ 5\,Gyr. The maximal difference is larger than the
typical uncertainty ($<$ 2\,Gyr) in stellar population studies (see
Fig. 9). The older the populations, the bigger the difference
between ages fitted via bsSSPs and ssSSPs, although the differences
between the Lick indices of old bsSSPs and ssSSPs are smaller (see
Section 3). The reason is that the differences among the Lick
indices of populations with different ages are much less for old
populations than for young populations (see Fig. 9 for comparison).
Therefore, the ss-fitted ages of galaxies can be much less than
bs-fitted ages, because most galaxies, especially early-type ones,
have old (7$\sim$8\,Gyr) populations and their metallicities are not
big (peak near 0.002) \citep{Gallazzi:2005}. However, ss-fitted
metallicities of populations are similar to bs-fitted values,
compared to the typical uncertainties ($\sim$ 0.002). Therefore, if
some stars are binaries, smaller ages will be measured via comparing
the observational H$\beta$ and [MgFe] indices of galaxies with those
of theoretical ssSSPs. This is more significant for metal-poor
stellar populations. In our testing populations, on average, the
ss-fitted metallicities are 0.0010 poorer than bs-fitted values,
while ssSSP fitted ages are younger than bs-fitted values, by
0.3\,Gyr for all and 1.8\,Gyr for old ($\geq$ 7\,Gyr) testing
populations. In this work, the ss-fitted stellar ages and
metallicities of testing bsSSPs are obtained by finding the best-fit
populations in a grid of theoretical populations with intervals of
stellar age and metallicity of 0.1\,Gyr and 0.0001, respectively. A
least square method is used in the fit. In addition, it is found
that the bs-fitted ages of populations can be calculated from the
ss-fitted ages and metallicities (with RMS of 1.45\,Gyr) via the
equation
\begin{equation}
t_{\rm b} = (0.17 + 8.27Z_{\rm s}) + (1.38 - 14.45Z_{\rm s})t_{\rm
s},
\end{equation}
where $t_{\rm b}$, $Z_{\rm s}$ and $t_{\rm s}$ are the bs-fitted
age, ss-fitted metallicity and age, respectively. The relation
between bs-fitted ages and ss-fitted stellar-population parameters
can be seen in Fig. 10.
\begin{figure} 
\includegraphics[angle=-90,width=160mm]{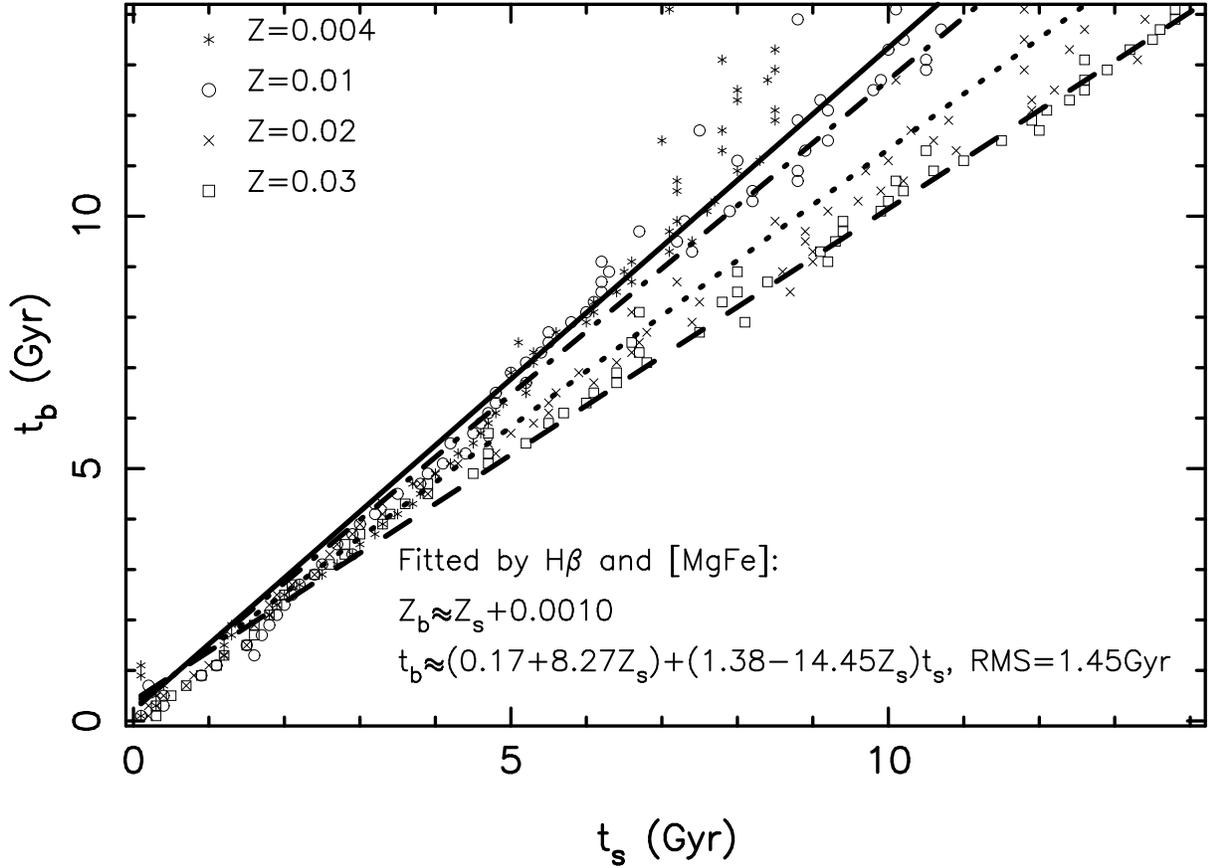}
\caption{The relation between bs-fitted ages of bsSSPs and their
ss-fitted stellar-population parameters, when using H$\beta$ and
[MgFe] \citep{Thomas:2003} to measure the ages and metallicities of
testing bsSSPs. Star, circle, cross, and square point represent
stellar populations with real metallicities of 0.004, 0.01, 0.02,
and 0.03, respectively. Solid, dash-dotted, dotted, and dashed lines
show the fitted relations between the bs-fitted ages and ss-fitted
results of populations, for ss-fitted metallicities of 0.003, 0.009,
0.019, and 0.029, respectively.}
\end{figure}
We find that ss-fitted ages are usually less than the bs-fitted ages
of populations, and the poorer the metallicity, the larger the
differences between the bs- and ss-fitted ages. Note that Eq. (2) is
not very accurate for metal-poor ($Z$ = 0.004) and old (age $>$
11\,Gyr) populations. The reason is that the H$\beta$ index
increases with age for metal-poor and old populations, while it
decreases with age for other populations.

\subsection{Photometric method}

In the photometric method, we fit stellar-population parameters
respectively via two pairs of colours, i.e., [$(u-R)$, $(R-K)$] and
[$(u-r)$, $(r-K)$]. The two pairs are shown to have the ability to
constrain the ages and metallicities of populations and can be used
to study the stellar populations of some distant galaxies (see
\citealt{Li:2008colourpairs}). The test shows that ss-fitted
metallicities are poorer than the bs-fitted metallicities of
populations. When taking [$(u-R)$, $(R-K)$] for this work, on
average, ss-fitted metallicities are 0.003 smaller than bs-fitted
values. It is 0.0035 when taking the pair [$(u-r)$, $(r-K)$]. The
distribution of a few testing bsSSPs in the $(u-R)$ versus $(R-K)$
grid of ssSSPs is shown in Fig. 11.
\begin{figure} 
\includegraphics[angle=-90,width=160mm]{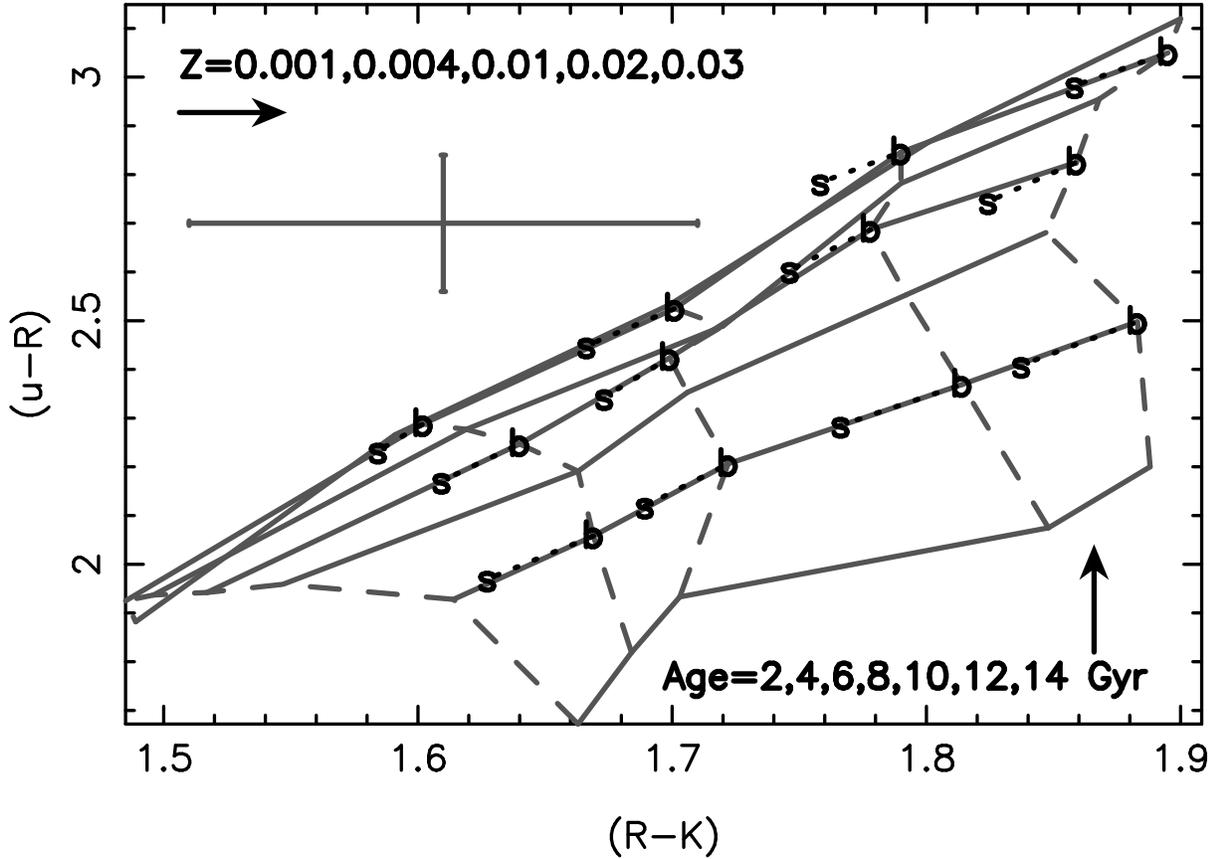}
\caption{Some bsSSPs on the $(u-R)$ versus $(R-K)$ grid of ssSSPs.
The position ``b'' shows the bs-fitted age and metallicity of a
bsSSP, and ``s'' shows the ss-fitted values. Dashed and solid lines
are for constant metallicity and constant age, respectively. Error
bars give typical observational uncertainties taken from the NED,
SDSS and 2MASS surveys.}
\end{figure}
In particular, it is found that the ss-fitted ages are correlated
with the bs-fitted ages of populations, which is independent of
metallicity. The relation (with a RMS of 0.72\,Gyr) between ss- and
bs-fitted ages of populations can be written as
\begin{equation}
t_{\rm b} = 0.24 + 0.93t_{\rm s},
\end{equation}
where $t_{\rm b}$ and $t_{\rm s}$ are bs- and ss-fitted ages,
respectively. It shows that the bs- and ss-fitted ages are similar.
The equation is clearly different from Eq. (2), because colours are
usually less sensitive to metallicity compared with metal-line Lick
indices. The relation between bs- and ss-fitted ages of populations
is shown in Fig. 12.
\begin{figure} 
\includegraphics[angle=-90,width=160mm]{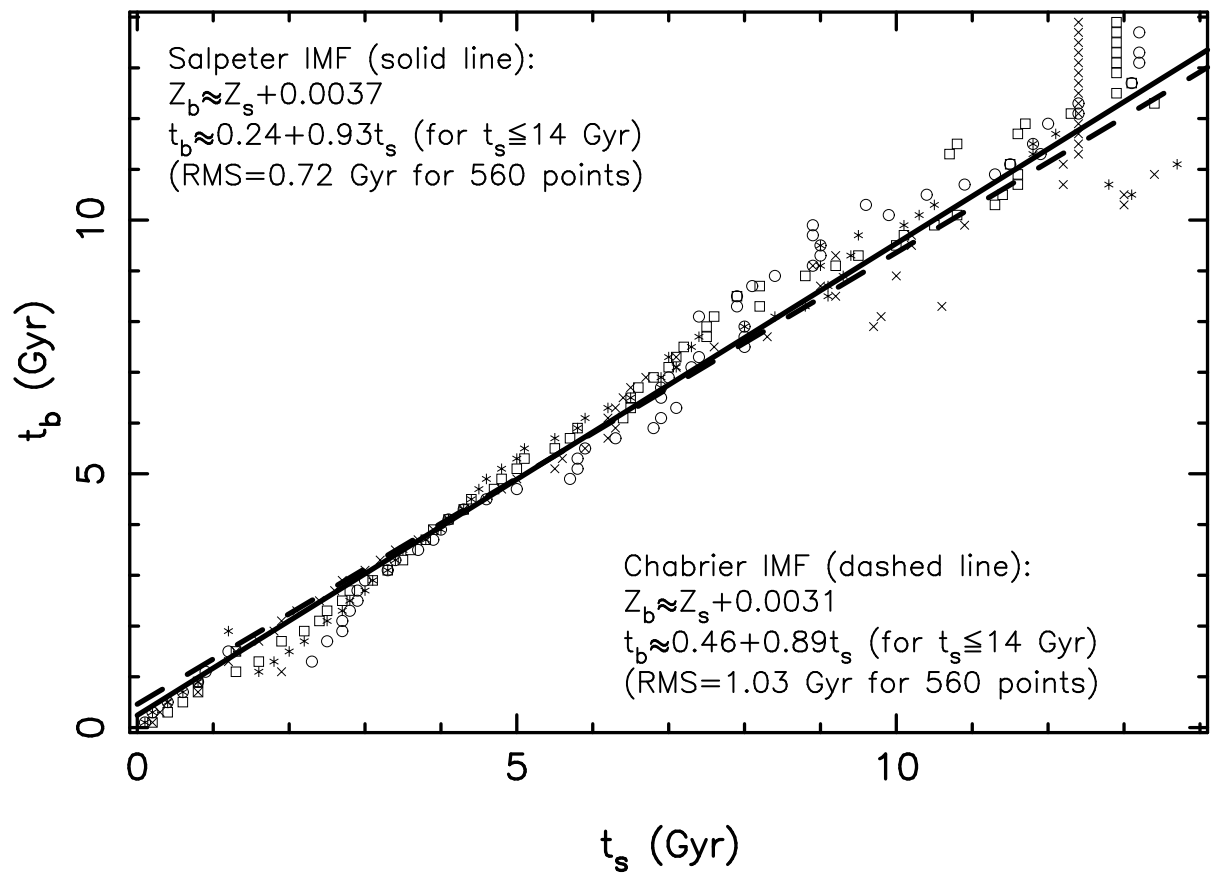}
\caption{The relation between bs- and ss-fitted ages and
metallicities of populations, when using $(u-R)$ and $(R-K)$ colours
to measure the parameters. Points are for populations with Salpeter
IMF and have the same meanings as in Fig. 10. Solid and dashed lines
show the fitted relations between bs- and ss-fitted ages, for
populations with Salpeter and Chabrier IMFs, respectively.}
\end{figure}
The figure shows the approximate relation between the bs- and
ss-fitted ages of populations, which is nearly independent of
metallicity. The equation is possibly useful to estimate the
absolute ages of distant galaxies and star clusters. Note that the
relation is presented for populations younger than 14\,Gyr, because
the age of the universe is shown to be smaller than about 14\,Gyr
\citep{WMAP:2003}.

When we use $(u-r)$ and $(r-K)$ colours to estimate the
stellar-population parameters of populations, we find that ss-fitted
metallicities are about 0.0035 smaller than the bs-fitted values.
The bs- and ss-fitted ages of populations can be approximately
transformed by
\begin{equation}
t_{\rm b} = 0.28 + 0.91t_{\rm s},
\end{equation}
where $t_{\rm b}$ and $t_{\rm s}$ are the bs- and ss-fitted ages,
respectively. The RMS of the fitted relation is 1.00\,Gyr. The
equation is similar to Eq. (3) but with larger scatter. This results
from the various metallicity and age sensitivities of the colours
used. One can see Fig. 13 for more details about the relation.
\begin{figure} 
\includegraphics[angle=-90,width=160mm]{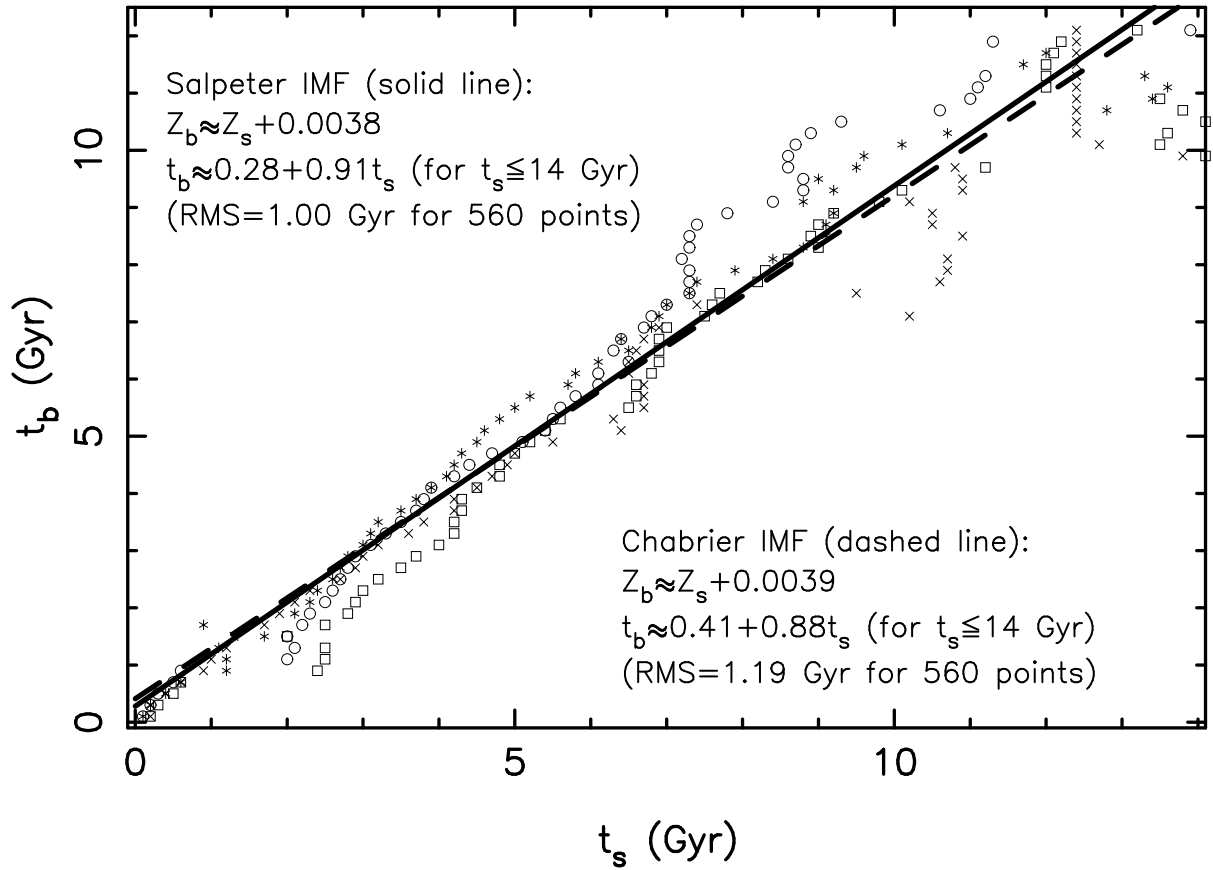}
\caption{Similar to Fig. 13, but for the results obtained via
$(u-r)$ and $(r-K)$ colours.}
\end{figure}

As a whole, from both the results obtained by Lick-index and
photometric methods, we are shown that bs-fitted stellar-population
parameters increase with ss-fitted ones. Therefore, using bsSSP
models instead of ssSSP models, similar results for relative studies
of stellar-population parameters of galaxies will be obtained.
However, if one wants to get the absolute stellar-population
parameters of galaxies and star clusters, the effects of binary
interactions should be taken into account, especially for metal-poor
populations. It can be conveniently done by taking the average
metallicity deviations and the relations between the bs-fitted ages
and ss-fitted results of populations, which were shown above.

\subsection{Results for populations with Chabrier initial mass function}

Some stellar populations with Salpeter IMF \citep{Salpeter:1955}
were taken as standard models for the work, but even if some
populations with other IMFs were taken instead, we can obtain
similar results. We have a test using populations with the Chabrier
IMF \citep{Chabrier:2003}. The result shows that ss-fitted
metallicities are 0.0011 less on average than bs-fitted results,
when taking H$\beta$ and [MgFe] for measuring stellar-population
parameters. The bs-fitted ages and ss-fitted stellar-population
parameters have a relation of $t_{\rm b} = (-0.06 + 20.63Z_{\rm s})
+ (1.46 - 18.76Z_{\rm s})t_{\rm s}$, where $Z_{\rm s}$ is the
ss-fitted metallicity, while $t_{\rm b}$ and $t_{\rm s}$ the bs- and
ss-fitted ages, respectively. When we used $(u-R)$ and $(R-K)$ to
estimate the stellar-population parameters of populations, ss-fitted
metallicities were shown to be 0.0031 smaller than bs-fitted values,
and bs-fitted ages can be calculated from ss-fitted results via
$t_{\rm b} = 0.46 + 0.89t_{\rm s}$. A similar relation for the
results fitted by $(u-r)$ and $(r-K)$ is $t_{\rm b} = 0.41 +
0.88t_{\rm s}$, with a deviation of 0.0039 in metallicity. As a
whole, the relations between bs- and ss-fitted results obtained via
populations with Salpeter and Chabrier IMFs are similar, compared to
the typical uncertainties of stellar-population parameter studies.
The comparisons of the results obtained via two IMFs can be seen in
Figs. 11 and 12.

\section{Discussions and Conclusions}
We investigated the effects of binary interactions on the
isochrones, SEDs, Lick indices and colours of simple stellar
populations, and then on the determination of light-weighted stellar
ages and metallicities. The results showed that binary interactions
can affect stellar population synthesis studies significantly. In
detail, binary interactions make stellar populations less luminous
and bluer, while making the H$\beta$ index larger and metal line
indices smaller compared to ssSSPs. The colour changes (2$\sim$5\%)
caused by binary interactions are smaller than the systematic errors
(about 6\%) of the $RPS$ model while similar to observational errors
(4$\sim$7\%). Note that the systematic error of 6\% did not take the
uncertainties due to the free parameters of the star model into
account (see Sect. 2). The changes (3$\sim$6\%) of Lick indices
caused by binary interactions are somewhat smaller than the
systematic errors (about 6\%) of stellar population synthesis model
but larger than observational errors (1$\sim$4\%). Therefore, if we
measure luminosity-weighted stellar-population parameters
(metallicity and age) via bsSSPs instead of ssSSPs, higher (0.0010
on average) metallicities and significantly larger ages will be
obtained via a Lick-index method, and significantly higher (about
0.0030) metallicities and similar ages will be obtained via a
photometric method. Because simple stellar population models are
usually used for studying the populations of early-type galaxies or
globular clusters, which possibly have old ($>$ 7\,Gyr) and relative
metal-poor populations, the changes ($\sim$ 1.8\,Gyr in age and
0.0030 in metallicity) caused by binary interactions in stellar ages
and metallicities are larger than the typical uncertainties. In
particular, we found that the relative results of stellar population
studies obtained by ssSSPs and bsSSPs are similar. The bs-fitted
stellar-population parameters can be calculated from the ss-fitted
ones, via equations presented by the paper.

The relations between bs-fitted ages and ss-fitted
stellar-population parameters are useful for some special
investigations. For example, when studying the age of the universe
via the stellar ages of some distant globular clusters, we can
estimate the absolute age of star clusters using ss-fitted results.
Although the results shown in this paper can help us to give some
estimates for the absolute stellar ages and metallicities of
galaxies, it is far from getting accurate values because of the
large uncertainties in stellar population models (see, e.g.,
\citealt{Yi:2003}). In addition, different stellar population models
usually give different absolute results for stellar population
studies. Note that the results obtained by the Lick-index method are
affected slightly by $\alpha$-enhancement, according to the work of
\citet{Thomas:2003}, but this is not the case for the results
obtained by photometric methods.

In this work, all bsSSPs contains about 50\% binaries with orbital
periods less than 100 yr (the typical value of the Galaxy). If the
binary (with orbital periods less than 100 yr) fraction of galaxies
are different from 50\%, the results shown in the paper will change.
The higher the fraction of binaries, the larger the difference
between ss- and bs-fitted stellar-population parameters. Thus the
results obtained by in this paper may not be proper for
investigating galaxies or star clusters with binary fractions
obviously different from 50\%. Furthermore, when building bsSSPs, we
assumed that the masses of the two components of a binary are
correlated \citep{Li:2007database}, according to previous works. We
did not try to take other distributions for secondary mass and
binary period in this work, because it is limited by our present
computing ability. We will give further studies in the future. The
differences between the Lick indices and colours of bsSSPs and
ssSSPs do not evolve smoothly with age. This possibly relates to the
method used to calculate the integrated features of stellar
populations. In fact, the Monte Carlo method usually leads to some
scatter (about 2\%) in the integrated features of populations. The
analytic fits and the binary algorithm used by Hurley code can also
lead to some scatter (about 5\%).

We investigated the effects of binary interactions via only some
simple stellar populations, but the real populations of galaxies and
star clusters are usually not so simple. In other words, the
populations of galaxies and star clusters seem to be composite
stellar populations including populations with various ages and
metallicities (e.g., \citealt{Yi:2005}). It seems that the effects
of binary interactions and population-mixing are degenerate. This is
a complicated subject, which requires further study.

\acknowledgments

We greatly acknowledge the anonymous referee for two constructive
referee's reports, Profs. Licai Deng, Tinggui Wan, Xu Kong, and
Xuefei Chen for useful discussions, and Dr. Richard Simon Pokorny
for greatly improving the English. This work is supported by the
Chinese National Science Foundation (Grant Nos 10433030, 10521001,
2007CB815406) and the Youth Foundation of Knowledge Innovation
Project of The Chinese Academy of Sciences (07ACX51001). 

\clearpage
\begin{thebibliography}{31}
\expandafter\ifx\csname
natexlab\endcsname\relax\def\natexlab#1{#1}\fi

\bibitem[{{Bruzual} \& {Charlot}(2003)}]{Bruzual:2003}
{Bruzual}, G., \& {Charlot}, S. 2003, {MNRAS}, 344, 1000

\bibitem[{{Chabrier}(2003)}]{Chabrier:2003}
{Chabrier}, G. 2003, {ApJ}, 586, L133

\bibitem[{{Davies} {et~al.} (2004)}]{Davies:2004}
{Davies}, M., {Piotto}, G., \& {De Angeli}, F. 2004, {MNRAS}, 349,
129

\bibitem[{{Duquennoy} \& {Mayor}(1991)}]{Duquennoy:1991}
{Duquennoy}, A., \& {Mayor}, M. 1991, {A\&A}, 248, 485

\bibitem[{{Fioc} \& {Rocca-Volmerange}(1997)}]{Fioc:1997}
{Fioc}, M., \& {Rocca-Volmerange}, B. 1997, {A\&A}, 326, 950

\bibitem[{{Gallazzi} {et al.}(2005)}]{Gallazzi:2005}
{Gallazzi}, A., {Charlot}, S., {Brinchmann}, J., {White}, S., 2005,
{MNRAS}, 362, 41

\bibitem[{{Goldberg} \& {Mazeh}(1994) {Goldberg} \& {Mazeh}}]{Goldberg:1994}
{Goldberg}, D.,  {Mazeh}, T.,  1994, {A\&A}, 282, 801

\bibitem[{{Han} {et~al.} (1995) {Han}, {Podsiadlowski} \& {Eggleton}}]{Han:1995}
{Han}, Z.,  {Podsiadlowski}, P.,    {Eggleton}, P.~P., 1995,
{MNRAS}, 272, 800

\bibitem[{{Han} {et~al.}(2007){Han}, {Podsiadlowski}, \&
  {Lynas-Gray}}]{Han:2007}
{Han}, Z., {Podsiadlowski}, P., \& {Lynas-Gray}, A.~E. 2007,
{MNRAS}, 380, 1098

\bibitem[{{Hurley} {et~al.}(2007){Hurley}, {Aarseth}, \& {Shara}}]{Hurley:2007}
{Hurley}, J.~R., {Aarseth}, S.~J., \& {Shara}, M.~M. 2007, {ApJ},
665, 707

\bibitem[{{Hurley} {et~al.}(2002){Hurley}, {Tout}, \& {Pols}}]{Hurley:2002}
{Hurley}, J.~R., {Tout}, C.~A., \& {Pols}, O.~R. 2002, {MNRAS}, 329,
897

\bibitem[{{Kroupa} {et~al.}(1993){Kroupa}, {Tout}, \& {Gilmore}}]{Kroupa:1993}
{Kroupa}, P., {Tout}, C.~A., \& {Gilmore}, G. 1993, {MNRAS}, 262,
545

\bibitem[{{Lejeune} {et~al.}(1998){Lejeune}, {Cuisinier}, \& {Buser}}]{Lejeune:1998}
{Lejeune}, T., {Cuisinier}, F., \& {Buser}, R. 1998, {A\&AS}, 130,
65

\bibitem[{{Li} \& {Han}(2007)}]{Li:2007database}
{Li}, Z., \& {Han}, Z. 2007, MNRAS, in press,
ArXiv:astro-ph/0708.1204

\bibitem[{{Li} \& {Han}(2008)}]{Li:2008colourpairs}
{Li}, Z., \& {Han}, Z. 2008, MNRAS, 385, 1270

\bibitem[{{Lodieu} {et~al.}(2007){Lodieu}, {Dobbie}, {Deacon}, {Hodgkin},
  {Hambly}, \& {Jameson}}]{Lodieu:2007}
{Lodieu}, N., {Dobbie}, P.~D., {Deacon}, N.~R., {Hodgkin}, S.~T.,
{Hambly},
  N.~C., \& {Jameson}, R.~F. 2007, {MNRAS}, 380, 712

\bibitem[{{Martins} {et~al.}(2005){Martins}, {Delgado}, {Leitherer},
  {Cervi{\~n}o}, \& {Hauschildt}}]{Martins:2005}
{Martins}, L.~P., {Delgado}, R.~M.~G., {Leitherer}, C.,
{Cervi{\~n}o}, M., \&
  {Hauschildt}, P. 2005, {MNRAS}, 358, 49

\bibitem[{{Mazeh} {et~al.}(1992) {Mazeh}, {Goldberg}, {Duquennoy} \&
  {Mayor}}]{Mazeh:1992}
{Mazeh}, T.,  {Goldberg}, D.,  {Duquennoy}, A.,    {Mayor}, M.,
1992, {ApJ}, 401, 265

\bibitem[{{Pinfield} {et~al.}(2003){Pinfield}, {Dobbie}, {Jameson}, {Steele},
  {Jones}, \& {Katsiyannis}}]{Pinfield:2003}
{Pinfield}, D.~J., {Dobbie}, P.~D., {Jameson}, R.~F., {Steele},
I.~A., {Jones},
  H.~R.~A., \& {Katsiyannis}, A.~C. 2003, {MNRAS}, 342, 1241

\bibitem[{{Pols} {et~al.}(1998){Pols}, {Hurley}, \& {Tout}}]{Pols:1998}
{Pols}, O., {Hurley}, J., \& {Tout}, C. 1998, in IAU Symposium, Vol.
191, IAU
  Symposium, 607

\bibitem[{{Salpeter}(1955)}]{Salpeter:1955}
{Salpeter}, E.~E. 1955, {ApJ}, 121, 161

\bibitem[{{Thomas} {et~al.}(2003){Thomas}, {Maraston}, \&
  {Bender}}]{Thomas:2003}
{Thomas}, D., {Maraston}, C., \& {Bender}, R. 2003, {MNRAS}, 343,
279

\bibitem[{{Thomas} {et~al.}(2005){Thomas}, {Maraston}, {Bender}, \& {Mendes de
  Oliveira}}]{Thomas:2005}
{Thomas}, D., {Maraston}, C., {Bender}, R., \& {Mendes de Oliveira},
C. 2005,
  {ApJ}, 621, 673

\bibitem[{{Tian} {et~al.}(2006){Tian}, {Deng}, {Han}, \& {Zhang}}]{Tian:2006}
{Tian}, B., {Deng}, L., {Han}, Z., \& {Zhang}, X.~B. 2006, {A\&A},
455, 247

\bibitem[{{Tout} {et~al.}(1997){Tout}, {Aarseth}, {Pols}, \&
  {Eggleton}}]{Tout:1997}
{Tout}, C.~A., {Aarseth}, S.~J., {Pols}, O.~R., \& {Eggleton}, P.~P.
1997,
  {MNRAS}, 291, 732

\bibitem[{{Vazdekis}(1999)}]{Vazdekis:1999}
{Vazdekis}, A. 1999, {ApJ}, 513, 224

\bibitem[{{Westera} {et~al.}(2002){Westera}, {Lejeune}, {Buser}, {Cuisinier},
  \& {Bruzual}}]{Westera:2002}
{Westera}, P., {Lejeune}, T., {Buser}, R., {Cuisinier}, F., \&
{Bruzual}, G.
  2002, {A\&A}, 381, 524

\bibitem[{{Bennett} {et~al.}(2003)}]{WMAP:2003}
{{Bennett}, C.~L., {Halpern}, M., {Hinshaw}, G., {Jarosik}, N.,
{Kogut}, A., {Limon}, M., {Meyer}, S.~S., {Page}, L., {Spergel},
D.~N., {Tucker}, G.~S., {Wollack}, E., {Wright}, E.~L., {Barnes}, C.
and {Greason}, M.~R., {Hill}, R.~S., {Komatsu}, E., {Nolta}, M.~R.
and {Odegard}, N., {Peiris}, H.~V., {Verde}, L., {Weiland}, J.~L.},
2006, {ApJS}, 148, 1

\bibitem[{{Worthey}(1994)}]{Worthey:1994}
{Worthey}, G. 1994, {ApJS}, 95, 107

\bibitem[{{Worthey} {et~al.}(1994){Worthey}, {Faber}, {Gonzalez}, \&
  {Burstein}}]{Worthey:1994lickdefinition}
{Worthey}, G., {Faber}, S.~M., {Gonzalez}, J.~J., \& {Burstein}, D.
1994,
  {ApJS}, 94, 687

\bibitem[{{Xin} {et~al.}(2007){Xin}, {Deng}, \& {Han}}]{Xin:2007}
{Xin}, Y., {Deng}, L., \& {Han}, Z.~W. 2007, {ApJ}, 660, 319

\bibitem[{{Yi}(2003)}]{Yi:2003}
{Yi}, S.~K. 2003, {ApJ}, 582, 202

\bibitem[{{Yi} {et~al.}(2005){Yi}, {Yoon}, {Kaviraj}, {Deharveng}, {Rich},
  {Salim}, {Boselli}, {Lee}, {Ree}, {Sohn}, {Rey}, {Lee}, {Rhee}, {Bianchi},
  {Byun}, {Donas}, {Friedman}, {Heckman}, {Jelinsky}, {Madore}, {Malina},
  {Martin}, {Milliard}, {Morrissey}, {Neff}, {Schiminovich}, {Siegmund},
  {Small}, {Szalay}, {Jee}, {Kim}, {Barlow}, {Forster}, {Welsh}, \&
  {Wyder}}]{Yi:2005}
{Yi}, S.~K., {Yoon}, S.-J., {Kaviraj}, S., {Deharveng}, J.-M.,
{Rich}, R.~M.,
  {Salim}, S., {Boselli}, A., {Lee}, Y.-W., {Ree}, C.~H., {Sohn}, Y.-J., {Rey},
  S.-C., {Lee}, J.-W., {Rhee}, J., {Bianchi}, L., {Byun}, Y.-I., {Donas}, J.,
  {Friedman}, P.~G., {Heckman}, T.~M., {Jelinsky}, P., {Madore}, B.~F.,
  {Malina}, R., {Martin}, D.~C., {Milliard}, B., {Morrissey}, P., {Neff}, S.,
  {Schiminovich}, D., {Siegmund}, O., {Small}, T., {Szalay}, A.~S., {Jee},
  M.~J., {Kim}, S.-W., {Barlow}, T., {Forster}, K., {Welsh}, B., \& {Wyder},
  T.~K. 2005, {ApJ}, 619, L111

\bibitem[{{Yungelson} \& {Tutukov}(1997)}]{Yungelson:1997}
{Yungelson}, L., \& {Tutukov}, A. 1997, in Advances in Stellar
Evolution, ed.
  R.~T. {Rood} \& A.~{Renzini}, 237

\bibitem[{{Zhang} {et~al.}(2004){Zhang}, {Han}, {Li}, \& {Hurley}}]{Zhang:2004}
{Zhang}, F., {Han}, Z., {Li}, L., \& {Hurley}, J.~R. 2004, {A\&A},
415, 117

\bibitem[{{Zhang} {et~al.}(2005){Zhang}, {Li}, \& {Han}}]{Zhang:2005}
{Zhang}, F., {Li}, L., \& {Han}, Z. 2005, {MNRAS}, 364, 503

\end{thebibliography}
\end{document}